\documentclass[preprint,showpacs,amsmath,amssymb]{revtex4}
\usepackage{graphicx}
\usepackage{color}
\begin{document}
\title{Effects of self-consistency violation in Hartree-Fock RPA calculations
for nuclear giant resonances revisited
}

\vspace{0.5cm}
\author{Tapas Sil$^1$, S. Shlomo$^1$, B. K. Agrawal$^{1,2}$, and P.-G. Reinhard$^3$}
\affiliation{
$^1$ Cyclotron Institute, Texas A\&M University, College Station, 
Texas 77843, USA\\
$^2$ Saha Institute of Nuclear Physics, 1/AF Bidhannagar, 
Kolkata 700064, India\\
$^3$ Institut f\"ur Theoretische Physik, Universit\"at Erlangen, 
D-91058 Erlangen, Germany,
}
\date{\today}

\begin{abstract}
We provide accurate assessments of the consequences of violations of 
self-consistency in Hartree-Fock (HF) based random phase 
approximation (RPA) calculations of the centroid energy $E_{cen}$ of isoscalar 
and isovector giant resonances of multi-polarities $L=0-3$ in a wide 
range of nuclei. This is done by carrying out highly accurate 
HF-RPA calculations neglecting the particle-hole
(ph) spin-orbit or Coulomb interaction in the RPA and comparing with 
the fully self-consistent HF-RPA results. We find that the shifts in the 
value of $E_{cen}$ due to self-consistency violation  associated
with the spin-orbit and Coulomb interactions are comparable or larger than 
the current experimental errors in $E_{cen}$. 
\end{abstract}
\maketitle
\section{Introduction}

The study of collective modes in nuclei provide very important information 
for understanding the structural and bulk properties of nuclear systems.
For example, 
the isovector giant dipole (IVGDR) mode is sensitive to
the symmetry energy, the centroid energy $E_{cen}$ of the 
isoscalar giant monopole resonance (ISGMR) 
is related to 
the value of the incompressibility modulus $K_\infty$ 
of symmetric nuclear matter 
\cite{boh1,lip1}, 
and low lying
collective states give a clue on the nuclear shell structure. 
These quantities are important ingredients not only for 
the description of finite nuclei but also for the study of heavy-ion 
collisions, supernovae and neutron stars. Recent developments in high 
precision experimental facilities make it possible to measure the 
centroid energy of the ISGMR with an error of $\delta E_{cen} \sim 0.1-0.3$ MeV 
\cite{you1,you2}. 
Using the approximate relation $(\delta K_\infty)/K_\infty=2(\delta
E_{cen})/E_{cen}$ and, for example, the recent experimental value
of $E_{cen}=13.96\pm0.20$ MeV for the ISGMR in 
$^{208}{\rm Pb}$, one has an error 
of $\delta K_\infty=6-9$ MeV for $K_\infty=200-300$ MeV.
This enhanced precision calls for a critical
accuracy check at the side of the the theoretical calculations with
the goal that the error in the calculated value of $E_{cen}$ is less
than the experimental error.  

The basic theory for the microscopic description of different modes of
giant resonances is the Hartree-Fock(HF) based random phase
approximation (RPA) \cite{boh2,shl1}.  
Although this
approach is conceptually well understood, actual calculations make
compromises for reasons of simplicity or numerical expense.  One can
hardly avoid limitations in the numerical basis and RPA phase
space. Furthermore, most of the presently available HF-RPA
calculations omit the painful to evaluate pieces of the RPA residual
interaction, namely its spin-orbit and/or Coulomb parts. We will call
that the self-consistency violation (SCV) in the following. It is
obvious that a very accurate calculation within HF-RPA demands a
sufficiently complete basis and in particular self-consistency, i.e.,
using exactly the same pieces in the residual interaction that have
been used in the underlying HF calculation.  Unfortunately, apart
from some fully self-consistent calculations
\cite{rei1,bla1,ter1,nak1,pie1}, most existing HF-RPA calculations are
contaminated by SCV. First assessments on the
effects of SCV on the strength functions of giant resonances were
obtained in Refs.  \cite{bka1,col1,bk2}. In Ref. \cite{bka1}, results
of elaborate studies of the effects of SCV on the constrained energy
($E_{con}$) and scaling energy ($E_s$) have been reported only for the
ISGMR. It is to be noted that the full self-consistent values of 
$E_{con}$ and $E_s$ were obtained using the constrained Hartree-Fock (CHF)
approach. 
It was pointed out \cite{bka1} that the SCV concerning spin-orbit and
Coulomb interactions may cause an error in $E_{con}$ of the ISGMR
which becomes larger than 1 MeV, i.e. as large as 5 times the
experimental error. This calls for a systematic assessment of the
effects of SCV on the excitation energies of other modes as well. 
Recently in Ref. \cite{ter1}, fully self-consistent calculations of
strength function were reported and the shift of the peak of the ISGMR
strength function due to SCV has been discussed for $^{120}{\rm
Sn}$. In Ref. \cite{nak1}, the effect of SCV has been assessed for the
IVGDR in $^{16}{\rm O}$ using the fully self-consistent approach of
small amplitude time dependent Hartree-Fock method.
We will continue here with discussing a greater variety of modes,
namely the sequence of isoscalar and isovector multi-pole resonances
in the range $L=0-3$ and a few examples from low lying collective
states.

Before attacking the main task, it is worthwhile to put the various
sources of uncertainties into perspective. The HF-RPA method optimizes
the modes in the space of one-particle-one-hole ($1ph$) excitations.
Correlations, i.e. $2ph$ and higher structures are not accounted for
explicitly. Such correlations have been very much discussed in the
past, for reviews see e.g. \cite{Ber83aR,Mah85aR,Rei94aR}. The main
effect is a collisional broadening of the strength distributions.
This can be accompanied by a certain shift of the resonance peak
position which grows with excitation energy and can reach orders of 1
MeV for the rather high lying isovector modes (in the range above 20
MeV).  However, the Skyrme forces employed in nuclear HF and RPA
calculations are effective forces which incorporate already a great
deal of correlations \cite{Ben03aR}. This reduces the correlation
effects on the peak positions \cite{Rei94aR,Gue93a}.  We adopt the
view that the net effect remains of order of a tenth of MeV for modes
with moderate excitation energy around and below 15 MeV.  A second
crucial aspect concerns limitations within the HF-RPA approach
itself. There are two major questions: the size of the RPA phase space
and the handling of the particle continuum. We take care to use a
sufficiently large expansion basis such that peak positions have
converged to uncertainties below 0.1 MeV. The artificial
discretization of the continuum has very little effect on the average
peak positions but limits the spectral resolution with which one can
reliably compute the strength distributions \cite{Rei05a}. 
We use a
large simulation box and properly adapted smoothing width.
After all, there remain the effects of SCV to be studied.
We demonstrate the accuracy of our fully self-consistent HF based RPA calculations
of the strength functions of giant resonances by comparing  (i) the RPA 
results with the corresponding ones of CHF for the
case of the ISGMR and (ii) the total energy weighted strengths with the 
corresponding energy weighted sum rules (EWSR).

\section{Formalities}

We calculate the strength function, 
\begin{equation}
S(E)= \sum_j E_j|<0|F_{L}|j>|^2\\
\label{ewg}
\end{equation}
following the fully self-consistent method based on 
$Q-P$ representation described in Ref. \cite{rei1,rei2}. 
In Eq. (\ref{ewg}), $|0>$ is the RPA ground state and the sum is over all 
RPA excited states $|j>$ with the corresponding excitation energy
$E_j$. The scattering operator $F_L$ is given by
\begin{eqnarray}
\label{issop}
F_L&=&\sum_i f(r_i)Y_{L0}(i),\,\,\,\,\qquad \qquad \qquad\qquad \qquad \qquad
\quad {\rm for\quad isoscalar},\\
F_L&=& \frac{Z}{A}\sum_n f(r_n))Y_{L0}(n) - 
\frac{N}{A}\sum_p f(r(p))Y_{L0}(p),\qquad {\rm for\quad isovector},
\label{ivsop}
\end{eqnarray}
with $f(r)=r^2,r^2$ and $r^3$ for monopole, quadrupole and octopole, 
respectively.
For the isovector dipole we have taken,  $f(r)=r$, whereas for the isoscalar 
dipole we adopt the scattering operator  $f(r)=r^3-\frac{5}{3}<r^2>r$
to eliminate the contribution of spurious state mixing \cite{bk2}.
Once we have the strength function, the energy moments can be calculated by,
\begin{equation}
\label{coneq3}
m_k=\int_0^\infty \omega^k S(\omega)d\omega.
\end{equation}
Then the centroid, constrained and scaling energies of the giant resonance 
are  computed  as,
\begin{equation}
\label{coneq4}
E_{cen}=\frac{m_1}{m_0},\,\,\,\,\,\,\,\;
E_{con}=\sqrt{\frac{m_1}{m_{-1}}},\,\,\,\,\,\,\,\;{\rm and}\;\,\,\,\,\,\,\,
E_{s}=\sqrt{\frac{m_3}{m_1}}.
\end{equation}

For the isoscalar ($T=0$) operator in Eq. (\ref{issop}), the EWSR 
is given by \cite{boh2,lip1}
\begin{equation}
m_1(L,T=0)=\frac{1}{4\pi}\frac{\hbar^2}{2m}A \left<g_L(r)\right>,
\label{ews}
\end{equation}
with 
\begin{equation}
\left<g_L(r)\right>=\frac{1}{A}\int g_L(r) \rho(r) 4\pi r^2 dr,
\label{gl1}
\end{equation}
where $\rho(r)$ is the HF ground state matter density distribution and  
\begin{equation}
\label{gl2}
g_L(r)=\left(\frac{df}{dr}\right)^2+L(L+1)\left(\frac{f}{r}\right)^2
.
\end{equation}
For the isovector ($T=1$) operator of Eq. (\ref{ivsop}), we have,
\begin{equation}
m_1(L,T=1)= \frac{NZ}{A^2} m_1(L,T=0)[1+\kappa-\kappa_{np}]
\label{ewv}
\end{equation}
where $\kappa$ is the enhancement factor due to the momentum dependence
of the effective nucleon-nucleon interaction, 
and is given by,
\begin{equation}
\label{enh_fac}
\kappa=\frac{(1/2)[t_1(1+x_1/2)+t_2(1+x_2/2)]}{(\hbar^2/2m)(4NZ/A^2)}
\frac{2\int g_L(r)\rho_p(r)\rho_n(r) 4\pi r^2 dr}{\int g_L(r)\rho(r) 4\pi r^2 dr}
,
\end{equation}
where $t_i$ and $x_i$ are the parameters of the Skyrme interaction.
The correction $\kappa_{np}$,  which arises
because of the difference in the profiles of the neutrons and protons
density distribution ( ie., 
since $\rho_n(r)-\rho_p(r) \ne \frac{N-Z}{A}\rho(r)$), is 
given by, 
\begin{equation}
\label{knp}
\kappa_{np}=\frac{(N-Z)}{A}\frac{A}{NZ}
\frac{\int g_L(r)[Z\rho_n(r)-N\rho_p(r)]4\pi r^2 dr}{\int g_L(r)\rho(r) 4\pi r^2 dr} 
.
\end{equation}

The fully self-consistent moment $m_{-1}$ for the ISGMR can also be calculated 
using the constrained Hartree-Fock  method \cite{boh1},
\begin{equation}
\label{coneq}
m_{-1}=\frac{1}{4\pi}\left.\frac{1}{2}\frac{d<r^2>_\lambda}{d\lambda}\right|_{\lambda=0}
=\frac{1}{4\pi}\left.\frac{1}{2}\frac{d^2E_\lambda}{d\lambda^2}\right|_{\lambda=0}
\end{equation}
where the mean square radius $<r^2>_\lambda$ and the energies $E_\lambda$ are
obtained using the solution of the constrained Hamiltonian 
$H_\lambda=H-\lambda r^2$. The self-consistent energy moment $m_1$
can be evaluated \cite{boh2} using the EWSR of 
Eqs. (\ref{ews})-(\ref{gl2}) with $f=r^2$,
\begin{equation}
\label{coneq1}
m_1=\frac{1}{4\pi}\frac{\hbar^2}{2m}4A<r^2>,
\end{equation}
where $m$ is the mass of the nucleon, $A$ is the mass number and $<r^2>$ is 
the mean square radius calculated using the  ground state HF wave function. 
Thus, using Eqs. (\ref{coneq}) and (\ref{coneq1}),
one can get the self-consistent 
value of $E_{con}$  which can also be used to 
check the accuracy of the RPA calculations (in particular, the
completeness of the RPA phase space).
For plotting purpose, we employ a Lorentzian smearing of the strength function
of Eq. (\ref{ewg}) obtaining,
\begin{equation}
\label{smear}
S(E)=\frac{1}{\pi}\sum_j
\frac{(\Gamma/2)|\langle 0|F_L|j\rangle|^2}{
(E_j-E)^2+(\Gamma/2)^2},
\end{equation}
where $\Gamma$ is the smearing parameter taken to be 2 MeV or larger.
Note, however, that for the evaluations of the 
energy moments, Eq. (\ref{coneq3}), and the energies of 
various giant resonances,
we use a very small value for $\Gamma$ ($<0.2$ MeV).

Since we shall be investigating the magnitude of errors in the calculation
of the observables for giant resonances due to the violation of the 
self-consistency, it is  necessary to pay attention 
to the accuracy of HF and RPA calculations. 
We have taken a box of
size 30 fm. This large computational box allows a spectral resolution
of the strength functions in the particle continuum of about 0.8 MeV
\cite{Rei05a} which is by far sufficient for our present investigation.
The mesh size is taken to be 0.3 fm. The flexible formulation of the RPA 
\cite{rei1} in the $Q$-$P$ representation allows one to include 1ph excitations 
up to very high energy, just by taking the detailed 1ph space only for 
transitions up to few major single particle shell \cite{rei2}.   
We have taken the detailed 1ph space up to 4  particle major shell and 
checked the convergence of the results by comparing with those obtained 
with 6 particle major shell. 
We have adopted the SGII \cite{sg2} interaction for the 
HF based RPA calculations and the Slater approximation has been used 
for the Coulomb exchange term, consistently both in the HF and RPA calculations.
For the CHF calculations, we have taken a box of 15 fm with mesh size 0.1
fm and calculated each derivative in Eq. (\ref{coneq}) using five
point formula with the increment 0.02 in the constraining parameter $\lambda$. 

\section{Results and discussion}
In order to check the accuracy of our CHF calculations for $m_{-1}$ in ISGMR, 
we have given the values for $^{40}$Ca and $^{208}$Pb
in Table \ref{tab0}, calculated from Eq. (\ref{coneq})
using the mean square radii, $m_{-1}(r)$, and the energy, $m_{-1}(e)$,
methods, for different 
values of the constraining parameter $\lambda$. For comparison,
the values of $m_{-1}$ obtained from RPA calculations Eqs. (\ref{ewg}) and 
(\ref{coneq3}) are given in the last 
column. It can be seen that $m_{-1}(r)$ differs from $m_{-1}(e)$ by 
0.7\% in the case of $\lambda=0.01$ for $^{40}$Ca. 
For $\lambda=0.02$, 
the deviations are very small (0.04\%) for both the nuclei. 
In the following, we will use the value of $\lambda=0.02$ in our calculations.
We add that for $\lambda=0.02$ the RPA
values of $m_{-1}$, obtained from Eq. (\ref{coneq3}) 
using the integration range 
$0-60$ MeV, compare very well with the CHF values (within 0.6\%). 
This clearly demonstrates the high accuracy of  
our CHF and HF-RPA calculations.

A very important necessary condition for a fully self-consistent HF based RPA
calculations for giant resonances is to obtain the full EWSR from the
calculated RPA strength function. In order to see how this necessary
condition is fulfilled in our present calculation,
we compare in the Table \ref{tab_ew}, the $m_1$ values calculated 
using the RPA strength functions in Eq. (\ref{coneq3}), ($m_1(RPA)$)
with the corresponding EWSR obtained from 
Eqs. (\ref{ews}) and (\ref{ewv}) 
for the nuclei $^{40}$Ca, $^{90}$Zr
and $^{208}$Pb for $L=0-3$ and $T=0$ and 1.
It is seen that for all of these nuclei and for all the modes,
the deviations of $m_1(RPA)$ from the corresponding EWSR
are very small  (less than 0.3\%). 
This once again shows the high accuracy of  our  HF-RPA calculations.
We note that for the SGII interaction we have 
$\kappa=0.314$, 0.381, 0.314 and 0.253 and the values of 
$\kappa_{np}=0.010$, 0.000, 0.010 and 0.024
for the isovector $L=0$, 1, 2 and 3 in $^{208}$Pb, respectively.
We point out that 
the correction term $\kappa_{np}$ in (\ref{ewv}) for isovector modes 
which is usually 
missing in the literature is not negligible for asymmetric nuclei. 
As an example, for $^{208}$Pb, $L=3$, 
$\kappa_{np}$ has an effect of 2\% in the calculation of EWSR. 
The effect of $\kappa_{np}$ will be more significant for nuclei near the 
drip lines because of the large difference between the neutron 
and proton density distributions 
 and also for the large asymmetry $(N-Z)/A$ (see Eq. (\ref{knp})). 

In Table \ref{tab1}, we have given the energies 
$E_{con}$, $E_s$ and $E_{cen}$ of the ISGMR
for a host of nuclei ranging from very light 
to heavy,  including some neutron or proton rich nuclei. The second column 
of Table \ref{tab1} indicates 
the kind of calculation --- SC, LS or CO. The symbol SC corresponds to the 
fully self-consistent calculation, i.e., both 
the spin-orbit and Coulomb interactions are present in HF as well as 
in RPA. The spin-orbit violation (LS) means that the spin-orbit and Coulomb 
interactions are present in HF but the ph 
spin-orbit interaction is missing in the 
RPA calculation.
Finally, the Coulomb violation CO means that the spin-orbit and Coulomb
interactions are present in HF but the ph 
Coulomb interaction is dropped out from 
the RPA calculations.  
The constrained energies $E_{con}$ calculated 
with the CHF approach using Eqs. (\ref{coneq}) and (\ref{coneq1}) 
are presented in the third column. 
The values of  $E_{con}$, $E_s$ and $E_{cen}$ of the ISGMR
calculated in HF based RPA, are given  
in the next three columns. 
The range of integration 
for calculating the ISGMR energies from the strength function 
(Eqs. (\ref{coneq3}) and (\ref{coneq4})) is 0-60 MeV. 
Comparison of $E_{con}$ obtained from  the fully self-consistent (SC) 
HF-RPA calculation of the strength function
(forth column) with those obtained from the constrained HF method 
(third column) shows 
a maximum difference of 0.06 MeV. 
This indicates that 
the accuracy of our RPA calculation is very good. It is to be noted that 
the effect of the SCV of LS and CO are similar in $E_{con}$ and $E_{cen}$,
but a little different in $E_s$. The shift of the centroid energy 
$\delta E_{cen}$ due to LS or CO violation are listed  in the last column. 
The effects of violation for the spin-orbit (LS)
interaction are very small for $^{16}{\rm O}$, $^{40,60}{\rm Ca}$ 
and $^{80,110}{\rm Zr}$
but robust in other nuclei, particularly for $^{100}{\rm Sn}$ 
($\sim$1.2 MeV) and 
$^{56}{\rm Ni}$ ($\sim$1.8 MeV). This can be understood because the effect of 
the omission of the ph spin-orbit interaction is
expected to show up prominently only for the spin unsaturated
nuclei like $^{56}{\rm Ni}$
and $^{100}{\rm Sn}$ here. Hence, for the spin closed    
$^{16}{\rm O}$, $^{40,60}{\rm Ca}$ and $^{80,110}{\rm Zr}$ 
nuclei, the effects of 
violation of spin-orbit interaction are very small.
On the other hand, the effect of the CO violation seems to be 
dependent on the position of the nuclei in the Segre chart. 

The energies of the ISGMR are obtained from the ratios of the moments $m_k$
in Eq. (\ref{coneq4}).  In Table \ref{tab01}, we give
for $^{40}{\rm Ca}$ and $^{208}{\rm Pb}$ 
the values of $m_1$, $m_0$ and $m_{-1}$
obtained from the fully self-consistent HF-RPA calculations (SC) along with 
those values obtained from calculations where self-consistency is violated 
due to the omission of ph spin-orbit (LS) and ph Coulomb (CO) interactions.
Comparison of the $m_1$ values calculated in 
fully self-consistent RPA with the EWSR
obtained from Eq. (\ref{coneq1}) (given in column 3) shows that the RPA 
values are smaller than the corresponding 
EWSR values  by less than 0.3\%. The violations 
of self-consistency (LS or CO) has little effect on values of $m_1$ (0.04\%) 
while the effects of SCV are considerably larger in $m_{-1}$ and $m_0$ which 
mainly cause the shift in the  giant resonance energies. This shows that 
obtaining a value of  
$m_1$ in a RPA calculation which is very close to the EWSR 
does not indicate an accurate and fully self-consistent
implementation of HF based RPA theory (see also \cite{bk2}).
  
In theoretical study of giant resonances, the evaluation of the strength 
function $S(E)$ is needed to  compute the energy moments (\ref{coneq3}) 
and hence the centroid energy, constrained energy, etc.  Therefore,
it is worth seeing how the strength function as such gets affected by the self
consistency violation.  In Fig. 1, we display 
the variation of the $S(E)$
with energy E for isoscalar excitations of different 
multi-polarities ($L=0-3$), for the 
nucleus ${}^{208}{\rm Pb}$ as a representative case. We have smoothened all the
strength functions by a Lorentzian with a smearing parameter $\Gamma=2$
MeV (see Eq. (\ref{smear})) in order to wipe out artefacts from the discretization of the
continuum and to provide a smooth curve for better comparison of the 
effects of SCV in $S(E)$. In the top panel of Fig. 1 we have three curves for 
the ISGMR (L=0,T=0) ---
the fully self-consistent (SC) (solid line) result, with violation for ph  spin-orbit 
interaction in RPA
(LS) (dashed line) and the result with violation of 
ph Coulomb interaction in RPA (CO) (open circle). 
All these three 
curves have similar single-peaked structure. 
But it is significant to notice that the violation of LS has 
pushed the peak to a higher energy by  almost by 0.7 MeV, whereas the 
position of the peak of the curve has moved toward the lower energy side  
by around 0.4 MeV due to the self-consistency violation for the Coulomb 
interaction. This implies that the self consistency violation may 
cause an uncertainty in the calculated value of 
the centroid energy  $E_{cen}$ of ISGMR for  ${}^{208}{\rm Pb}$
which is twice in magnitude compared to the experimental one . 
This uncertainty, 
as we have discussed above, would impose an error of about 20 MeV in 
the prediction of the value of $K_\infty$. 
Since  the effects of LS and CO violations are 
opposite to each other, there would be a partial  cancellation if one 
does not take both of the ph LS and CO interactions simultaneously in 
the RPA calculations. 
For the isoscalar giant dipole (L=1,T=0) resonance (ISGDR), 
the effects of neglecting the 
Coulomb and spin-orbit terms in the 
ph interaction are similar to those for  ISGMR. 
Violation 
of LS leads to a push of  the peak of $S(E)$ toward a higher energy whereas the 
violation of CO  pulls the peak to a lower energy. 
But the magnitude of the
shift of the peak is much smaller in comparison with the ISGMR. Specifically, 
the ph Coulomb interaction
shows very little  effect.
We display similar curves for the higher multi-polarities, L=2 and 3, in the 
lower two panels. A shift of the position of the peak of $S(E)$ is 
observed once we 
neglect either the ph Coulomb or spin-orbit interactions 
in the RPA calculation. In both cases of
the quadrupole and octopole giant resonances, 
the effect of the violation of self consistency
acts in the same direction which is in contrary to the ISGMR and ISGDR cases.
Therefore, the total shift in the peak energy of $S(E)$ will be very 
significant if one omits both Coulomb and spin-orbit ph interactions 
in the RPA calculations.     
Note also the significant shifts in the energies and strengths of the 
low lying isoscalar $L=1-3$ states due to SCV. 

In Fig. 2, we plot the strength 
functions  for the isovector modes for $^{208}{\rm Pb}$ 
in the same way as we have done in Fig. 1 for
the isoscalar mode.  Here we have used in the calculations the same 
parameters as those 
taken for the isoscalar modes except for the smearing parameter $\Gamma=10$ MeV
for the isovector giant monopole resonance (IVGMR). 
It is clear from the figure that in the isovector modes  (for all considered 
multi-polarities)  the strength functions are almost insensitive to the self 
consistency violation  due to the omission of  the 
ph spin-orbit  interaction in the RPA calculation. 
On the other hand, the absence of the ph Coulomb
interaction in the RPA calculation pushes the strength 
function significantly toward 
lower energy and this is most prominent in the isovector dipole mode where 
the shift is around 0.4 MeV. 

In Table \ref{tab2},
we present, for a wide range of nuclei, the self consistent 
centroid energies $E_{cen}(SC)$ and
their shifts $\delta E_{LS}$ and $\delta E_{CO}$ due to the 
self-consistency violations for ph spin-orbit and Coulomb interactions
in the RPA calculations, respectively. We have given results 
for different multi-polarities
($L=0,1,2$ and 3) for both the isoscalar and the isovector modes.
It is evident from the table that for isoscalar modes, the effect
of LS or CO violations are most prominent for monopole resonance
which are almost 2 to 3 times larger than those for the other 
multi-polarities. 
Note that if we drop the ph LS and CO interactions simultaneously in 
the RPA calculations
for $^{208}{\rm Pb}$ ISGMR, the shift of the centroid energy is 
$\delta E_{LS,CO}=-0.30$ MeV which is comparable  to  the experimental  
uncertainty. The effects of the LS and CO self-consistency violations 
are found to be somewhat smaller
for $L=2$ and 3 in comparison with the  ISGMR for all of these nuclei. But for
the $L=2$ and 3 modes,
$\delta E_{LS}$ and $\delta E_{CO}$ are found to be of the same sign. Therefore,
their combined effect on the centroid energy is  significant.
The SCV associated with the Coloumb interaction 
(in the RPA level) affects $E_{cen}$ considerably in the dipole 
and monopole modes of the isovector channel.

Finally, we have checked the effect of SCV on low lying (isoscalar)
collective states.  Test cases were the lowest $3^-$ in $^{90}$Zr and
the lowest $2^+$ in $^{208}{\rm Pb}$.  The low lying states are much more
sensitive to the size of the RPA phase space. We have used here an
expansion basis up to about 2000 MeV to ensure sufficient convergence.
The results are shown in Table \ref{tab3}. The effects are of the same
order as found for the giant resonances. However, for these low lying
states a mismatch of about 0.5 MeV is large compared to the total
excitation energy. A fully consistent calculation is compulsory for
studying spectra in that energy range.

\section{Conclusion}

In summary, we have carried out highly accurate fully self-consistent
Hartree-Fock (HF) based random phase approximation (RPA) calculations 
for the strength functions of isoscalar and isovector 
$L=0-3$ modes in a wide range of nuclei.
We have quantified very accurately the effects of 
self-consistency violations in the calculations of the energies 
of giant resonances of nuclei within the HF based RPA. 
We have studied the cases of SCV due to the omission of the 
spin-orbit (LS) or/and Coulomb (CO) ph interactions 
and mainly focus on their effects on the  centroid energy $E_{cen}$.
Here we 
consider both isoscalar and isovector modes of multi-polarities $L=0-3$.
It is found, for the wide range of nuclei considered here, that the effects of 
violations of self-consistency due to the ph 
LS or CO interactions are  most significant for the 
ISGMR. For the ISGMR, the absence of the ph LS  interaction 
tends to increase $E_{cen}$, whereas the violation due to 
ph CO interaction decreases $E_{cen}$.
For the spin unsaturated nuclei (such as 
${}^{56}{\rm Ni}$ and  ${}^{100}{\rm Sn}$), the shift in 
$E_{cen}$ is robust ($\sim 1.5$ MeV) which is almost 5 times larger 
than the experimental uncertainty. 
For other higher multi-polarities, the individual effect of the ph 
LS and CO interactions
are somewhat smaller than those for the 
 ISGMR. But for the quadrupole and octopole 
modes, the LS and CO self-consistency violations
both tends to reduce the centroid energy. Hence, the effect 
of SCV on $E_{cen}$ in these modes are significant ($0.3-0.6$ MeV) if 
one neglects the ph spin-orbit and Coulomb interactions simultaneously in 
the RPA calculation.

\acknowledgments{This work was supported in part by the US National 
Science Foundation under Grant No. PHY-0355200 and the US Department of Energy
under the Grant No. DOE-FG03-93ER40773.}

%



\newpage
\begin{table}
\caption{ The values of $m_{-1}$ of $^{40}$Ca and $^{208}$Pb for ISGMR obtained 
from the  mean square radii, $m_{-1}(r)$, and energy, $m_{-1}(e)$ 
methods (Eq. (\ref{coneq})), are presented for different values of the increment of the 
constraining parameter $\lambda$.
For comparison, the RPA values are given in the last column.}
\begin{center}
\begin{tabular}{|c|ccc|c|} \hline \hline
\multicolumn{1}{|c|}{Nucleus}&
\multicolumn{3}{|c|}{CHF}&
\multicolumn{1}{|c|}{RPA}\\
\multicolumn{1}{|c|}{}&
\multicolumn{1}{|c}{$\quad \lambda\quad$}&
\multicolumn{1}{c}{$\quad\quad m_{-1}(r)\quad$}&
\multicolumn{1}{c|}{$\quad\quad m_{-1}(e)\quad$}&
\multicolumn{1}{|c|}{$\quad\quad m_{-1}\quad$}\\
   & (fm$^{-2}$ MeV) & (fm$^4$ MeV$^{-1}$) 
   & (fm$^4$ MeV$^{-1}$) & (fm$^4$ MeV$^{-1}$) \\
\hline
${}^{40}{\rm Ca}$  & 0.01 & 6.512 & 6.467 & \\    
                   & 0.02 & 6.469 & 6.466 &6.426 \\
                   & 0.03 & 6.466 & 6.463 & \\    
 \hline
${}^{208}{\rm Pb}$ & 0.01 & 230.01 & 230.57 & \\    
                   & 0.02 & 230.57 & 230.50 & 230.83 \\ 
                   & 0.03 & 230.27 & 229.89 & \\    
 \hline
 \end{tabular}
\end{center}
\label{tab0}
\end{table}
\begin{table}
\caption{Comparison of $m_1$ calculated from RPA strength function S(E), 
$m_1(RPA)$ (Eq. (\ref{coneq3}) integrated up to $E_{max}\sim 100$ MeV),  
with those obtained from the 
energy weighted sum rules (EWSR) in Eqs. (\ref{ews}) and (\ref{ewv}). 
The ratio $R=m_1(RPA)/EWSR$ indicates that the maximum deviation 
of $m_1(RPA)$ from EWSR in this table is less than 0.3\%.
}
\begin{center}
\begin{tabular}{|c|c|cc|cc|cc|}\hline \hline
\multicolumn{2}{|c|}{Mode}&
\multicolumn{2}{|c|}{Ca40}&
\multicolumn{2}{|c|}{Zr90}&
\multicolumn{2}{|c|}{Pb208}\\
\hline
\multicolumn{1}{|c|}{L}&
\multicolumn{1}{|c|}{T}&
\multicolumn{1}{|c}{EWSR}&
\multicolumn{1}{c|}{R}&
\multicolumn{1}{|c}{EWSR}&
\multicolumn{1}{c|}{R}&
\multicolumn{1}{|c}{EWSR}&
\multicolumn{1}{c|}{R}\\
\hline
0 & 0 & 2889 & 0.9983 & 10505 & 0.9994 & 41872 & 0.9971\\
  & 1 & 896.3 & 0.9992 & 3330 & 0.9994 & 13041 & 0.9998\\
\hline
1 & 0 & 57253 & 0.9990 & 289907 & 0.9995 & 1823110 & 0.9999\\
  & 1 & 64.62 & 0.9999 & 148.9 & 1.0001 & 337.8 & 0.9999\\
\hline
2 & 0 & 7222 & 1.0001 & 26262 & 1.0007 & 104681 & 1.0015\\
  & 1 & 2241 & 1.0000 & 8326 & 0.9998 & 32604 & 1.0002\\
\hline
3 & 0 & 238240 & 0.9996 & 1300645 & 0.9999 & 8584813 & 1.0003\\
  & 1 & 69328 & 0.9994 & 389266 & 0.9996 & 2519881 & 1.0001\\
\hline
\end{tabular}
\end{center}
\label{tab_ew}
\end{table}
\renewcommand{\baselinestretch}{1.0}
\begin{table}
\caption{The constrained ($E_{con}$), scaling ($E_{s}$) and the centroid 
energies ($E_{cen}$), in MeV, of the ISGMR of various nuclei are given for 
fully self-consistent HF-RPA calculations (SC) along with those 
obtained for the calculations where the self-consistency is violated 
due to the neglecting of the ph spin-orbit (LS) and Coulomb (CO) 
interactions in RPA. We adopt the range $(\omega_1-\omega_2)=(0-60)$ MeV 
for the energy integration. The effects
of the SCV on the centroid energies are given in the last column. 
We have also given in Column 3, the  constrained energies obtained 
from fully self-consistent constrained Hartree-Fock (CHF) calculations.  
The Skyrme interaction SGII \cite{sg2} was used.
}
\begin{center}
\begin{tabular}{|c|c|c|cccc|} \hline \hline
\multicolumn{1}{|c|}{Nucleus}&
\multicolumn{1}{|c|}{$\quad \theta_{v}\quad$}&
\multicolumn{1}{|c|}{CHF}&
\multicolumn{4}{|c|}{RPA}\\
\multicolumn{1}{|c|}{}&
\multicolumn{1}{|c|}{}&
\multicolumn{1}{|c|}{$\quad E_{con}\quad$}&
\multicolumn{1}{c}{$\quad\quad E_{con}\quad$}&
\multicolumn{1}{c}{$\quad E_{s}\quad$}&
\multicolumn{1}{c}{$\quad E_{cen}\quad$}&
\multicolumn{1}{c|}{$\quad \delta E_{cen}\quad $}\\
\hline
${}^{16}{\rm O}$  & SC & 24.025 & 23.985 & 26.846 & 24.584 & \\
            & LS &        & 23.710 & 26.905 & 24.384 & 0.200\\
            & CO &        & 23.890 & 26.722 & 24.481 & 0.103\\
 \hline
${}^{40}{\rm Ca}$ & SC & 21.131 & 21.186 & 22.516 & 21.469 & \\
            & LS &        & 21.026 & 22.605 & 21.362 & 0.107\\
            & CO &        & 20.960 & 22.258 & 21.234 & 0.235\\
 \hline
${}^{60}{\rm Ca}$ & SC & 16.203 & 16.154 & 18.804 & 16.948 & \\
            & LS &        & 15.956 & 18.886 & 16.812 & 0.136\\
            & CO &        & 16.093 & 18.660 & 16.865 & 0.083\\
 \hline
${}^{56}{\rm Ni}$ & SC & 20.177 & 20.179 & 20.864 & 20.311 & \\
            & LS &        & 21.853 & 23.056 & 22.075 & -1.764\\
            & CO &        & 19.883 & 20.550 & 20.009 & 0.302\\
 \hline
${}^{80}{\rm Zr}$ & SC & 17.841 & 17.873 & 18.629 & 18.050 & \\
            & LS &        & 17.842 & 18.752 & 18.051 & -0.001\\
            & CO &        & 17.527 & 18.267 & 17.699 & 0.351\\
 \hline
${}^{90}{\rm Zr}$ & SC & 17.814 & 17.832 & 18.261 & 17.914 & \\
            & LS &        & 18.512 & 19.233 & 18.645 & -0.731\\
            & CO &        & 17.532 & 17.944 & 17.609 & 0.305\\
 \hline
${}^{110}{\rm Zr}$ & SC & 15.294 & 15.248 & 16.410 & 15.585 & \\
             & LS &        & 15.207 & 16.596 & 15.584 & 0.001\\
             & CO &        & 15.080 & 16.160 & 15.392 & 0.193\\
 \hline
${}^{100}{\rm Sn}$ & SC & 17.156 & 17.207 & 17.645 & 17.282 & \\
             & LS &        & 18.351 & 19.086 & 18.488 & -1.206\\
             & CO &        & 16.730 & 17.189 & 16.804 & 0.478\\
 \hline
${}^{116}{\rm Sn}$ & SC & 16.261 & 16.288 & 16.832 & 16.400 & \\
             & LS &        & 16.868 & 17.659 & 17.027 & -0.627\\
             & CO &        & 15.961 & 16.478 & 16.062 & 0.338\\
 \hline
${}^{144}{\rm Sm}$ & SC & 15.276 & 15.305 & 15.806 & 15.372 & \\
             & LS &        & 15.991 & 16.740 & 16.100 & -0.728\\
             & CO &        & 14.919 & 15.459 & 14.987 & 0.385\\
 \hline
${}^{208}{\rm Pb}$ & SC & 13.475 & 13.454 & 13.910 & 13.522 & \\
             & LS &        & 14.080 & 14.767 & 14.196 & -0.674\\
             & CO &        & 13.079 & 13.545 & 13.142 & 0.380\\
 \hline
 \end{tabular}
\end{center}
\label{tab1}
\end{table}
\begin{table}
\caption{ The values of $m_k$ of $^{40}$Ca and $^{208}$Pb for ISGMR obtained
from fully self-consistent RPA calculations (SC) along with those
obtained from the calculations where the self-consistency is violated
due to the neglecting of the ph spin-orbit (LS) and Coulomb (CO)
interactions in RPA. The EWSR values of  $m_1$ obtained from Eq. (\ref{coneq1})
are given in column 3.
}
\begin{center}
\begin{tabular}{|c|c|c|ccc|} \hline \hline
\multicolumn{1}{|c|}{Nucleus}&
\multicolumn{1}{|c|}{$\quad \theta_{v}\quad$}&
\multicolumn{1}{|c|}{EWSR}&
\multicolumn{3}{|c|}{RPA}\\
\multicolumn{1}{|c|}{}&
\multicolumn{1}{|c|}{}&
\multicolumn{1}{|c|}{$m_1$(Eq. (\ref{coneq1}))}&
\multicolumn{1}{|c}{$m_1$}&
\multicolumn{1}{c}{$\quad\quad m_{-1}\quad$}&
\multicolumn{1}{c|}{$m_0$}\\
& & (fm$^4$ MeV$^{-1}$) & (fm$^4$ MeV) & (fm$^4$ MeV$^{-1}$) & (fm$^4$)\\
\hline
${}^{40}{\rm Ca}$ & SC & 2888.7 & 2884.4 & 6.426 & 134.35\\    
                         & LS &  & 2884.6 & 6.525 & 135.03\\      
                         & CO &  & 2885.9 & 6.569 & 135.91\\
 \hline
${}^{208}{\rm Pb}$ & SC & 41872.5 & 41783.7 & 230.83 & 3090.03\\
                          & LS & & 41782.6 & 210.76 & 2943.37\\
                          & CO & & 41819.6 & 244.49 & 3182.06\\
 \hline
 \end{tabular}
\end{center}
\label{tab01}
\end{table}
\newpage
\renewcommand{\baselinestretch}{1.4}
\begin{table}
\caption{Fully self-consistent HF-RPA results of 
centroid energies ($E_{cen}(SC)$) and their shifts 
(in MeV) due to self-consistency violations via spin-orbit 
($\delta E_{LS}=E_{cen}(SC)-E_{cen}(LS)$) 
and Coulomb ($\delta E_{CO}=E_{cen}(SC)-E_{cen}(CO)$) 
ph interactions are presented 
for different modes ($L=0-3$, $T=0,1$) of some nuclei of experimental interest.
The ranges of energies of integrations ($\omega_1-\omega_2$) 
for the calculations of the centroid energies
are given in  columns 3 and 7. The Skyrme interaction SGII \cite{sg2} was
used.
}
\begin{center}
\begin{tabular}{|c|c|cccc|cccc|} \hline \hline
\multicolumn{1}{|c|}{Nucleus}&
\multicolumn{1}{|c|}{\quad L \quad }&
\multicolumn{4}{|c|}{Isoscalar}&
\multicolumn{4}{|c|}{Isovector}\\
\multicolumn{1}{|c|}{}&
\multicolumn{1}{|c|}{}&
\multicolumn{1}{c}{\quad $\omega_1-\omega_2$ \quad}&
\multicolumn{1}{c}{\quad $E_{cen}(SC)$ \quad }&
\multicolumn{1}{c}{\quad $\delta E_{LS}$ \quad }&
\multicolumn{1}{c|}{\quad $\delta E_{CO}$ \quad }&
\multicolumn{1}{c}{\quad $\omega_1-\omega_2$ \quad }&
\multicolumn{1}{c}{\quad $E_{cen}(SC)$ \quad }&
\multicolumn{1}{c}{\quad $\delta E_{LS}$ \quad }&
\multicolumn{1}{c|}{$\quad \delta E_{CO}$ \quad }\\
 \hline
${}^{40}{\rm Ca}$ & 0 &  5-40 & 21.31 & 0.12 & 0.21& 10-60& 
                           32.87 & 0.05 & 0.17\\
& 1 &  22-60 & 32.60 & -0.09 & 0.08 & 5-40 &
                           19.03 & -0.01 & 0.18\\
& 2 &  8-26  & 17.16 & 0.24 & 0.07 & 10-60 &
                           30.46 & -0.01& 0.07\\
& 3 &  22-45 & 31.48 & 0.30 & 0.02 & 24-65 &
                           39.87 & 0.02 & 0.02\\
 \hline
${}^{56}{\rm Ni}$ & 0 &  5-40 & 20.20 & -1.68 & 0.30& 10-60& 
                           33.74 & -0.29 & 0.18\\
& 1 &  22-60 & 32.78 & -0.54 & 0.10 & 5-40 &
                           19.01 & -0.16 & 0.21\\
& 2 &  8-26  & 17.52 & 0.32 & 0.12 & 10-60 &
                           30.68 & -0.02 & 0.10\\
& 3 &  22-45 & 31.17 & 0.55 & 0.03 & 24-65 &
                           40.40 & 0.14 & 0.04\\
 \hline
${}^{90}{\rm Zr}$ & 0 &  5-35 & 17.89 & -0.68 & 0.30 & 10-55 &
                           31.77 & -0.10 & 0.27\\
& 1 &  18-50 & 28.88 & -0.33 & 0.09 & 5-35 &
                           16.78 & -0.03 & 0.29\\
& 2 &  8-20 & 14.64 & 0.15 & 0.17 & 10-50 &
                           27.53 & -0.05 & 0.14\\
& 3 &  20-40 & 27.31 & 0.29 & 0.05 & 22-60 &
                           36.52 & -0.01 & 0.05\\
 \hline
${}^{116}{\rm Sn}$ & 0 &  5-35 & 16.38 & -0.59 & 0.34 & 10-50 &
                           30.26 & -0.20 & 0.30\\
& 1 &  18-45 & 27.39 & -0.33 & 0.10 & 5-35 &
                           15.75 & -0.08 & 0.32\\
& 2 &  8-18 & 13.60 & 0.14 & 0.19 & 10-45 &
                           25.82 & -0.17 & 0.16\\
& 3 &  18-32 & 25.37 & 0.40 & 0.05 & 22-55 &
                           34.87 & -0.10 & 0.05\\
 \hline
${}^{144}{\rm Sm}$ & 0 &  5-35 & 15.34 & -0.69 & 0.39 & 10-50 &
                           29.92 & -0.08 & 0.37\\
& 1 &  18-45 & 26.42 & -0.33 & 0.13 & 5-35 &
                           15.02 & -0.03 & 0.39\\
& 2 &  8-18 & 12.97 & 0.11 & 0.28 & 10-45 &
                           24.75 & -0.06 & 0.22\\
& 3 &  18-32 & 23.95 & 0.14 & 0.04 & 20-50 &
                           33.23 & -0.04 & 0.10\\
\hline
${}^{208}{\rm Pb}$ & 0 &  5-30 & 13.50 & -0.66 & 0.38 & 10-45 &
                           27.79 & -0.08 & 0.39\\
& 1 &  16-40 & 24.04 & -0.35 & 0.12 & 5-30 &
                           13.66 & -0.02 & 0.43\\
& 2 &  8-16 & 11.64 & 0.11 & 0.19 & 10-40 &
                           22.64 & -0.05 & 0.25\\
& 3 &  15-28 & 21.13 & 0.24 & 0.04 & 20-50 &
                           30.92 & -0.07 & 0.08\\
 \hline
 \end{tabular}
\end{center}
\label{tab2}
\end{table}

\begin{table}
\caption{The effect of SCV for low lying isoscalar states in two selected
cases. The Skyrme interaction SGII \cite{sg2} was used.
}
\begin{center}
\begin{tabular}{|c|c|cccc|} \hline \hline
\multicolumn{1}{|c|}{Nucleus}&
\multicolumn{1}{|c|}{\quad L \quad}&
\multicolumn{4}{|c|}{energies [MeV]}\\
& &\quad  SC \quad & \quad LS \quad &\quad CO \quad &\quad LS-CO \quad \\
 \hline
${}^{90}{\rm Zr}$ & 3 & 2.10 & 2.57 & 1.95 & 2.47 \\
\hline
${}^{208}{\rm Pb}$ & 2 &4.48 & 5.06 & 4.38 & 4.99 \\
 \hline
 \end{tabular}
\end{center}
\label{tab3}
\end{table}

\newpage
\begin{figure}
\includegraphics[width=0.85\linewidth, angle=0, clip=true]{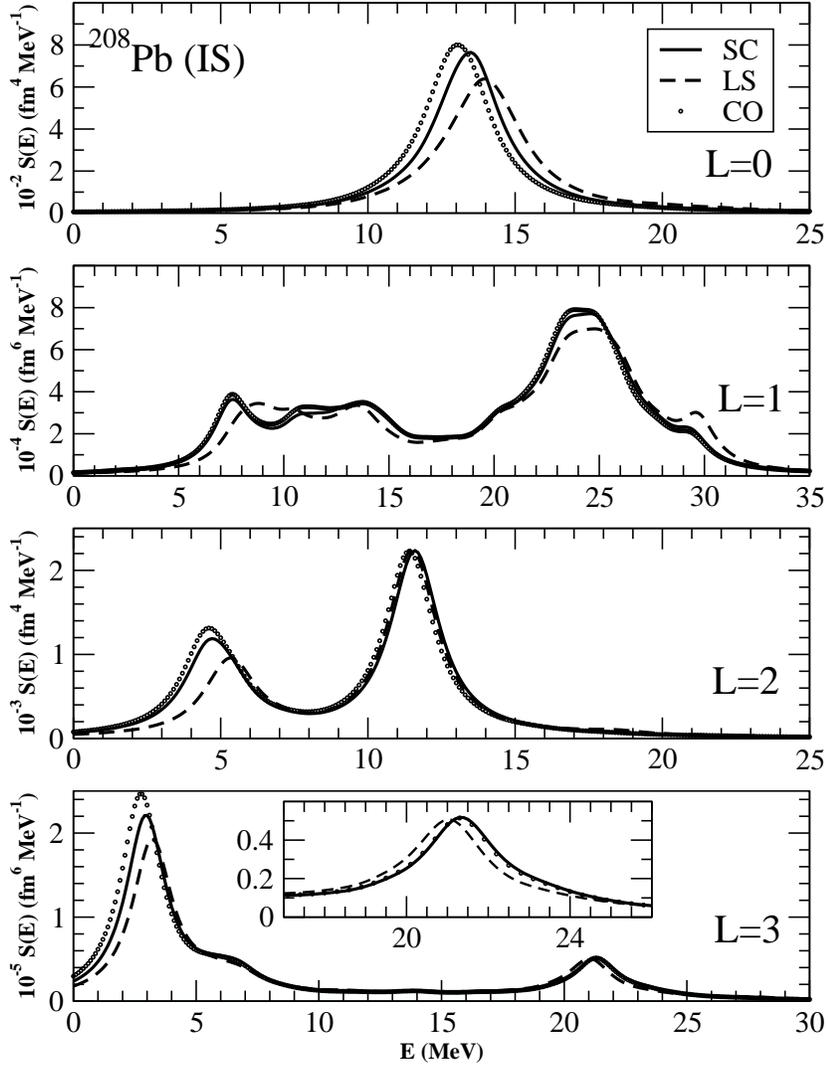}
\caption{\label{Fig1}HF-RPA results for 
isoscalar strength functions of $^{208}{\rm Pb}$ for
$L=0-3$ multi-polarities are displayed. SC (full line) corresponds to 
the fully self-consistent calculation where LS (dashed line) and CO (open 
circle) represent the calculations without the ph spin-orbit and Coulomb 
interactions in the RPA, respectively. A magnified giant resonance is 
shown in the inset (lowest panel) for $L=3$. 
The Skyrme interaction SGII \cite{sg2} was used.
}
\end{figure}
\newpage
\begin{figure}
\includegraphics[width=0.85\linewidth, angle=0, clip=true]{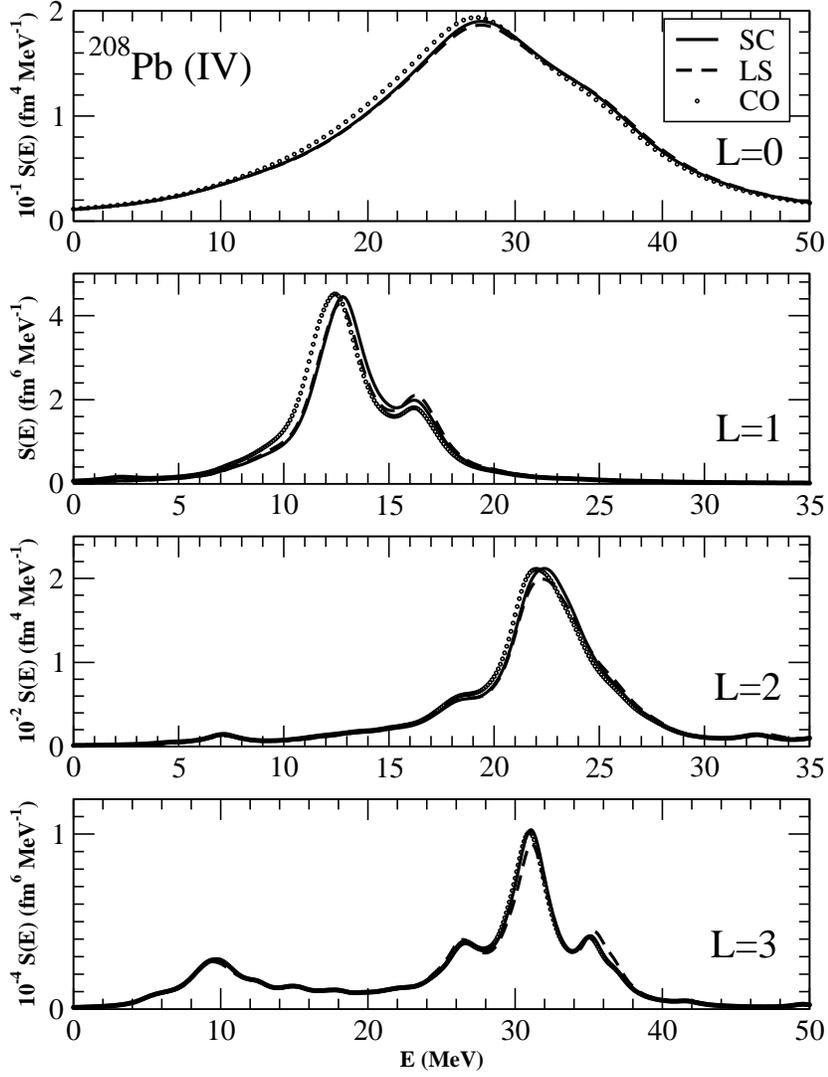}
\caption{\label{Fig2}Same as Fig.\ref{Fig1} but for isovector strength 
functions. }
\end{figure}
\end{document}